# Miniature fluorescence sensor for quantitative detection of brain tumour

Jean Pierre Ndabakuranye [a], James Belcourt [b], Deepak Sharma [a,c,d], Cathal D. O'Connell [a, e], Victor Mondal [f], Sanjay K. Srivastava [c,d], Alastair Stacey [b,g], Sam Long [h], Bobbi Fleiss [f], Arman Ahnood [a]

[a] School of Engineering, RMIT University, VIC 3000, Australia.
[b] School of Science, RMIT University, VIC 3000, Australia.
[c] Photovoltaic Metrology Section, Advanced Materials and Device Metrology Division, CSIR-National Physical Laboratory, New Delhi, 110012, India.
[d] Academy of Scientific and Innovative Research (AcSIR), Ghaziabad-201002, India.
[e] Aikenhead Centre for Medical Discovery, St Vincent's Hospital Melbourne, VIC 3065, Australia.
[f] School of Health and Biomedical Sciences, RMIT University, VIC 3000, Australia.
[g] Princeton Plasma Physics Laboratory, Princeton University, Princeton, 08540 New Jersey, USA
[h] Veterinary Referral Hospital, Victoria, Australia.

Fluorescence-guided surgery has emerged as a vital tool for tumour resection procedures. As well as intraoperative tumour visualisation, 5-ALA-induced PpIX provides an avenue for quantitative tumour identification based on ratiometric fluorescence measurement. To this end, fluorescence imaging and fibre-based probes have enabled more precise demarcation between the cancerous and healthy tissues. These sensing approaches, which rely on collecting the fluorescence light from the tumour resection site and its "remote" spectral sensing, introduce challenges associated with optical losses. In this work, we demonstrate the viability of tumour detection at the resection site using a miniature fluorescence measurement system. Unlike the current bulky systems, which necessitate remote measurement, we have adopted a millimetre-sized spectral sensor chip for quantitative fluorescence measurements. A reliable measurement at the resection site requires a stable optical window between the tissue and the optoelectronic system. This is achieved using an antifouling diamond window, which provides stable optical transparency. The system achieved a sensitivity of 92.3 % and specificity of 98.3 % in detecting a surrogate tumour at a resolution of $1\times1$ mm$^2$. As well as addressing losses associated with collecting and coupling fluorescence light in the current 'remote' sensing approaches, the small size of the system introduced in this work paves the way for its direct integration with the tumour resection tools with the aim of more accurate interoperative tumour identification.

1. Introduction

Successful tumour resection requires its precise identification to completely remove cancerous tissue while preserving the maximum amount of healthy tissue (Brown et al. 2016; Prickett and Ramsey 2017; Sanai and Berger 2018). This becomes even more critical when dealing with a complex part of the human body, such as the brain. Excessively removing healthy tissue may result in postoperative neurological impairment (Al-Mefty et al. 1990; Brown et al. 2017; Guerrini et al. 2022; Surma-aho et al. 2001; Zetterling et al. 2020). The proximity and sometimes diffuse nature of tumours require novel tools to better distinguish between healthy and cancerous tissue intraoperatively (Duffau 2005; Sanai and Berger 2018). Fluorescence-guided surgery (FGS) is one of the possible techniques that can address this (Bravo et al. 2017; Lauwerends et al. 2021; Nguyen and Tsien 2013; Richter et al. 2017; Stummer et al. 2006). In 1998, Stummer et al. (1998) discovered that when 5-aminolevulinic acid (5-ALA) is exposed to C6 glioma cells, it is metabolised and accumulates as protoporphyrin IX (PpIX) at a higher rate compared to healthy tissue (Stummer et al. 1998). In the current clinical practices, endogenous PpIX is induced in the tumour by administering the 5-ALA drug (Zhao et al. 2013). Exciting the tumour using blue light (usually at 405 nm) results in a pink PpIX glow (peak emission of approximately 635 nm). FGS has been used by surgeons to visualise tumours intraoperatively (Schupper et al. 2021). This approach has improved tumour resection outcomes by helping surgeons distinguish between healthy and cancerous tissue (Schupper et al. 2021; Wei et al. 2019). Although hugely valuable, given the inherently non-quantitative nature of visual tumour identification, this places a great emphasis on interoperative histopathological quantitative markers to guide the surgery (Valdes et al. 2011).

The density of cancerous cells (malignancy) directly correlates with the fluorescence strength (Zhao et al. 2013). Leveraging on the correlation between intensity and tumour malignancy, fluorescence measurement has been used to quantify tumour presence (Bendsoe et al. 2007; Zhao et al. 2013). Beyond gross visual localisation of the tumour, this principle provides an avenue for quantitative assessment of the surgical site with the aim of distinguishing more precisely between cancerous and healthy tissues (Valdés et al. 2011). To this end, interoperative fluorescence imaging has been used to quantitatively assess the fluorescence strength and identify its spatial distribution (Belykh et al. 2016; Lauwerends et al. 2021; Reichert et al. 2021; Valdes et al. 2014; Valdes et al. 2019; Yang et al. 2003). Another clinical approach which has been explored is the use of sensing probes in combination with FGS. These probes use a bundle of optical fibres housed within a stainless-steel probe to deliver excitation light to the tissue and collect the tissue's fluorescence (Haj-Hosseini et al. 2010; Lakomkin and Hadjipanayis 2019; Richter et al. 2017). These methods have relied on the use of conventional spectrometers to measure the tissue fluorescence spectrum. The critical feature of spectrometers is their high spectral resolution (typically below 5 nm) enabled using diffraction optics. However, this type of optical technology also means that spectrometers are typically bulky instruments. Although there has been great progress towards their miniaturisation (Malinen et al. 2014; Toulouse et al. 2021), nevertheless, the spectral measurements have thus far been performed 'remotely' and away from the resection site. This introduces several optical loss pathways that make the detection of low intensity fluorescence emission challenging. Although this is typically countered by increasing the excitation light intensity, this approach is not ideal as it limits the time window during which FGS can be performed, owing to the increased PpIX photobleaching rate at high excitation intensities (Belykh et al. 2018; Scott et al. 2000).

This work introduces tumour fluorescence detection using a miniature spectral sensing chip. The sensor achieves spectral measurement using an array of colour filters shown in Figure 1 (a), forgoing the need for diffraction gratings that are used in conventional spectrometers. Indeed similar approach has been demonstrated in earlier works (Dandin et al. 2007; Wang et al. 2007). This work builds on these by using the filter-based sensor for quantitative tumour detection. The spectral sensor used here incorporates a CMOS based photodetector and electronic circuitry combined with polymer based optical filters. As highlighted in Figure 1 (a), this approach results in orders of magnitude size reduction compared to conventional spectrometers. This enables the miniature spectral sensing system to be placed in close contact with the resection site. The small size of the sensor chip makes it possible to integrate it with the excitation source, additional optical filters and an antifouling diamond window onto a single printed circuit board as illustrated in the system schematic in Figure 1 (b) and the corresponding image in Figure 1 (c). We investigate the viability of using miniature spectral sensing for quantitative assessment of PpIX in brain tumour surrogate to evaluate capability of the sensing system to distinguish between fluorescence signatures of healthy and tumour tissues (Figure 1 (d) and (e)). Forgoing the need for any optical fibres or imaging systems, the small footprint of system shown in Figure 1 (c) facilitated the direct mapping of the brain tissue's fluorescence at low illumination intensities. The newly developed sensing system is used to directly collect the emitted light from the *ex-vivo* brain and quantitatively map the spatial distribution of PpIX fluorescence. Statistical analyses are used to compare the sensing system map shown in Figure 1 (f), with the actual region with elevated PpIX shown in Figure 1 (g) to determine the system's selectivity and sensitivity at 1 mm$^2$ spot sizes.

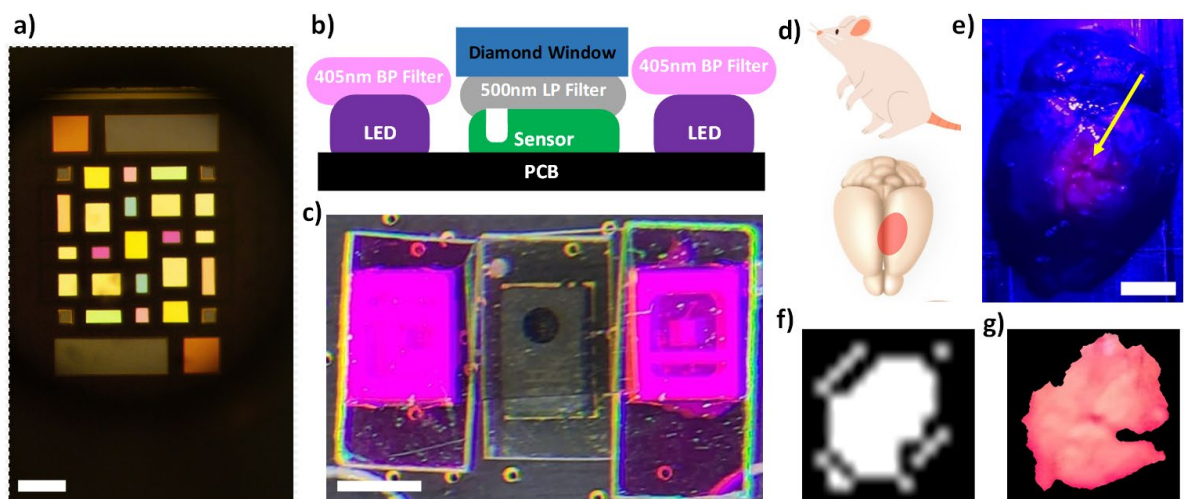

**Figure 1: Overview of miniature fluorescence sensing system used in quantitative mapping of rat brain. (a) Multispectral sensor chip which utilised an array of filters for spectral sensing (150 μm scale bar). (b) Cross sectional schematics and (c) top view image of the miniature fluorescence sensing system, consisting of the multispectral sensor chip, two LED excitation sources, bandpass (BP) and low pass (LP) optical filters and antifouling diamond window (2 mm scale bar). (d)  *ex-vivo* rat brain showing the approximate location of tumour mimicking PpIX induced fluorescence (e) Fluorescence image of brain with the arrow pointing to**

**area of PpIX induced tumour surrogate (5 mm scale bar). (f) Region of the brain with pronounced tumour fluorescence signature mapped using miniature fluorescence sensing system, compared to (g) mapped using fluorescence imaging.**

The use of pigment colour filters approach results in poorer spectral resolution compared with diffraction-based spectrometers. A typical CMOS integrated pigment colour filter offers measurement across several broad filter windows, with typical a Full Width at Half Maximum (FWHM) of 40~50 nm (see SI Fig. 1). This is an order of magnitude larger than conventional spectrometers which offer spectral resolutions that are less than 5 nm. Whilst this may appear as a drawback, given the clear spectral distinction between the autofluorescence and PpIX fluorescence, the use of conventional spectrometers in this application may be redundant. This is because a typical autofluorescence peak is centred at 510 nm with an FWHM of 118 nm (482 - 600 nm), whilst PpIX exhibits a 635 nm peak emission with an FWHM of 14 nm (627 nm - 641 nm) (Croce et al. 2003; Kruijt et al. 2008; Mehidine et al. 2019), resulting in a minimal spectral overlap between the two emissions. Beyond wider spectral bands, polymer based optical filters provide a poorer ability in blocking light outside their transmission window (low optical density), compared to spectrometers. From the application perspective this means the sensor chip's PpIX fluorescence measurement channel would be contaminated by the light generated due to tissue autofluorescence and vice versa. Nevertheless, the sensor chip's optical density is sufficient, given the large PpIX fluorescence / autofluorescence contrast exhibited between different types of tumours (Ishihara et al. 2007; Kaneko and Kaneko 2016). This is because a typical contrast error using polymer based optical filters is less than 0.01, significantly smaller than the contrast required to distinguish between healthy (0.7) and various types of tumours (1.3 to 30.9) (Kaneko and Kaneko 2016). As well as examining these hypotheses, this work investigates the capability of sensor chip in collecting dim light intensities – a vital attribute for successful FGS application.

Existing systems use optical fibres or imaging systems to collect the fluorescence emissions. Optical fibres lose a significant portion of the light at their interfaces. Similarly, efficient collection of diffused emissions using an imagining system is challenging. These losses create a challenge in the probing of low-intensity fluorescence emissions. Whilst our proposed approach addresses some of the challenges associated with light collection and coupling losses due to the use of fibre or imaging systems, it necessitates an optical window at the interface between the sensor chip and the tissue. In addition to optical transparency, minimal background fluorescence, mechanical robustness, and the requisite biocompatibility/sterilizability, the optical window must exhibit antifouling properties (Li et al. 2023; Sunny et al. 2016). This is to ensure it maintains a consistent optical property. Adherence of healthy/tumour tissue or blood to the surface of the optical window would result in erroneous fluorescence measurement. To ensure the viability of this approach, we introduced a diamond based antifouling window, as depicted in the schematic in Figure 1 c). As well as addressing the key requirements, hydrogen-terminated diamond is hydrophilic with water contact angle of 75~81º (Kaibara et al. 2003; Mertens et al. 2016). Moreover, earlier works have demonstrated that diamond offers a low friction coefficient (De Barros Bouchet et al. 2012; Hayward 1991; Konicek et al. 2008; Sutton et al. 2013). The hydrophobic surface helps various fluids (blood, saline) slide off the window's surface, whilst low friction minimises tissue adhesion. A combination of these attributes may address the requirements for an antifouling surgical window. In this work, we investigate the viability of a diamond optical window for surgical applications, by characterising its optical performance and *in-vitro* tribological properties.

2. Experimental methods

    2.1. Sensing approach

The measured fluorescence spectrum of PpIX is shown in SI Fig. 2. The spectral characteristics of healthy brain tissue and typical PpIX-containing brain tumours have spectral peaks at 510 nm and 635 nm, respectively (Haj-Hosseini et al. 2010). Based on these spectral signatures, a quantitative approach has been established whereby a diagnostic contrast parameter (Ratio) can be obtained from equation (1) which correlates with tumour malignancy (Bendsoe et al. 2007).

$$Ratio = \frac{I_{635\,nm} - I_{background}}{I_{510\,nm}} \quad (1)$$

In equation (1) PpIX peak fluorescence intensity at 635 nm and the brain's peak autofluorescence intensity at 510 nm are used. $I_{background}$ corresponds to the brain's background autofluorescence intensity at 635 nm and calculated using peak fitting to multiwavelength spectral measurements (Haj-Hosseini et al. 2010; Stummer et al. 1998).

In this work, we have adopted a similar approach. We have used emission intensities at 635 nm and 514 nm, as depicted in Equation (*2*). However, we have neglected the $I_{background}$ as our proposed system does not provide the information needed to determine it. In the equation (1) $I_{background}$ is needed to calculate the ratio $I_{background}/I_{514\,nm}$. This ratio relates to the shape of the autofluorescence spectrum. Given that the autofluorescence from the tumour and healthy brain tissues are similar, we have assumed that the ratio of $I_{background}/I_{514\,nm}$ is constant. This can be replaced by a fixed offset α in equation (2). Here α = $I_{background}/I_{514\,nm}$ and it is assumed that it is constant offset.

$$Ratio = \frac{I_{635\,nm}}{I_{514\,nm}} - \alpha \quad (2)$$

    2.2. Ex-vivo animal brain tissue

Brain tissue was collected immediately post-mortem from a healthy adult female Wistar rat purchased from the Australian Bioresource Center (Australia) and held and handled under the Australian Rules for the care and guidance of animals and approved by the RMIT Animal Ethics Committee (AEC1933). A tumour surrogate region was created by microneedle injecting exogenous PpIX solution into the surface of a healthy brain at the centre of the somatosensory cortex over an area of ~ 5 mm² - Figure 1. A key consideration here is the difference in the fluorescence emission spectrum emitted by the exogenous PpIX compared to the 5-ALA induced endogenous PpIX. Earlier works have demonstrated comparable spectral characteristics (Haj-Hosseini et al. 2010), which is consistent with results in SI Fig. 2. PpIX solutions were first prepared by dissolving protoporphyrin IX disodium salt in methanol (Erkkilä et al. 2019; Lehtonen et al. 2022). PpIX disodium salt and methanol were procured from Sigma Aldrich (Castle Hill, NSW 2154). To avoid induced photobleaching and improve stability, PpIX solutions were kept in the dark at 5 degrees Celsius and used within seven days of preparation. In our study, the concentration ranges of PpIX solutions were matched to the *in-vivo* concentration ranges of low to high grade gliomas. Studies by Johansson et al. (2010) had estimated the maximum concentration of PpIX in a high grade glioma as 28.2 μmol/L which corresponds to ~ 15870 ng/mL using a molar mass of 562 g/L. Erkkilä, *et al*., reported a concentration range of 156 to 2500 ng/mL from low to high grade (Erkkilä et al. 2019). Our study used a concentration of 10000 ng/mL.

### 2.3. Brain optical phantom

Brain optical phantoms were used to further characterise the sensor's ability to detect the tumour and tumour margin. The focus was to mimic the brain tissue's fluorescence properties without alluding to tissue absorption and scattering. Phantoms are prepared by gently mixing 10 mL of room temperature water and 2 g of gelatine powder (McKenzie's Foods, Altona, VIC 3018) in a glass vial to create a semi-homogenous solution. The vial is sealed and placed in a 60 °C water bath for complete melting. The vial is removed from the water bath, and the gelatine solution is further gently diluted with warm water to achieve a 4 g/dL homogenous gel-like solution. The gelatine solution exhibited no detectable fluorescence when excited by 405 nm light source. Two fluorescent inks were used to mimic the brain tissue autofluorescence and PpIX fluorescence spectrum. An estimate of ~10 ml of green or pink inks were mixed with two separate gelatine aliquots with gentle mixing to avoid introducing air bubbles. The samples exhibited autofluorescence peaks at 514 and 625 nm, which mimic the brain autofluorescence and PpIX fluorescence, respectively – see SI Fig. 3 for the spectral distribution. This is consistent with earlier reported works (Birriel and King 2018) where the spectral fluorescence for various highlighter inks were reported.

The optical phantom of a 'tumour' surrounded by a 'healthy" tissue was created by placing a tumour phantom on a ~ 2 mm$^2$ brain phantom. This allowed some diffusion driven mixing between the tumour and brain phantom to create a tumour margin. The sample was placed onto a glass slide and transferred into a refrigerator at 5 degrees Celsius for complete sample solidification. The two inks were mixed to produce a ratio value of around 5 at the centre of the tumour in line with earlier works (Richter et al. 2017).

### 2.4. Optical components and systems

The spectral fluorescence measurements were carried out using a spectrometer (USB2000, Ocean Insight, Inc.) and a fibre optic cable (R400-7-VIS-NIR, Ocean Insight, Inc.). Spectral data capturing by the spectrometer was achieved using OceanView software. Optical power measurements were performed using a FieldMax II Laser Power Meter (Coherent Inc.). 405 nm LEDs (MPN: 153283407A212, Wurth Elektronik) were used as an excitation source. In addition to its peak emission at 405 nm, the LED exhibited a broad secondary peak centred at

~ 550 nm, which overlaps with the healthy brain's autofluorescence spectrum. This was addressed using a 405 nm bandpass filter (PN: 15-117, Edmund Optics) to reduce the secondary spectral shoulder from the excitation source - SI Fig. 4. Moreover, the high intensity of excitation to emission ratio and the use of filter-based approach for wavelength detection makes it necessary to use a filter to block the excitation light at the detector. This was achieved using a 425 nm long pass filter (PN: 84-736, Edmund Optics). Fluorescence imaging was performed under a 405 nm LED excitation source, using a conventional CCD camera equipped with a long pass filter.

### 2.5. Sensor chip

A miniature sensor chip (AS7343, ams OSRAM) was adopted in this work. The sensor chip offers 14 measurement channels. Of these channels, two channels centred at 514 nm and 635 nm were used in the measurement. The sensor chip circuitry incorporates 16-bit analogue-to-digital converters (ADC), intermediate frequency (IF) filters and I2C peripheral. These features enable minimal electrical noise from the adjacent high power LED coupling into the measurement. Key optical and electrical characteristics of the sensor are summarised in SI Table 1.

### 2.6. Diamond window

The diamond window consisted of an optical grade, single-crystal diamond (Element Six), with one side scaif polished, and the other side resin-bond polished. The sample was 4.5mm x 4.5mm x 0.3mm in size. The window was hydrogen terminated using an ASTeX microwave plasma-enhanced chemical vapour deposition (MW-PECVD) chamber whereby it was exposed to a hydrogen plasma for several minutes. The sample stage was heated to 800 Celsius when the plasma was ignited and then gradually increased from 400 W MW power and 10 Torr to 1.2 kW MW power and 75 Torr over 2.5 minutes. The plasma was then maintained at 800 Celsius stage temperature, 1.2 kW MW power, and 75 Torr chamber pressure for 5 minutes. The MW power was then gradually turned off over 2 minutes where the sample was then allowed to cool under hydrogen gas flow to ambient temperatures. Transmittance measurements were carried out using a white LED (Colour temperature: 6000K, forward voltage: 2.7 - 3.5 V, family: OSLON® SSL 150, part number: GWCSHPM1.EM, manufacturer: OSRAM).

The adhesion of brain tissue to the diamond and glass substrates was characterised using an Anton Paar MCR302 rheometer. The substrate of interest (diamond or glass) was mounted onto a 25 mm diameter parallel plate (PP25) measuring geometry and brought into contact with brain tissue, which was fixed onto the stage. A shear strain ramp was performed from a strain of 0.001 to 10 in oscillatory mode at a frequency of 1 rad/s. The measured stress values were normalised based on the area of the substrate (2.5 mm x 2.5 mm in each case). The yield stress was then calculated using the elastic stress method (Walls et al. 2003; Yang et al. 1986). All measurements were performed at 37 Celsius. Temporary adhesive (Crystal bond 509) was used throughout to attach the diamond window.

3. Results and discussions

   *3.1. Tumour and margin detection*

The proposed approach is centred on the viability of detecting the fluorescence signatures emitted from the tumour, tumour margin and healthy tissue using the sensor chip. This was investigated by testing tumour models and healthy brain constructed using an ex-vivo brain tissue (see section 2.2) and an optical phantom model (see section 0). While the *ex-vivo* animal brain model includes the complexity of the real brain tissue, facilitating a more realistic assessment of the newly developed sensing device, the phantom model selected for this work lacks the complexity of scattering and absorption properties of the actual brain tissue, making it ideally a simpler model to gain a better understanding of the performance of the newly developed setup from the fluorescence sensing perspective.

Figure 2 (a) shows a fluorescence image of the rat brain with the PpIX induced fluorescence region - the yellow dotted line indicates the margin. Figure 2 (b) shows the sensor system ratio readout, calculated using equation (*2*), as a function of displacement at steps of 0.5 mm measured along the boxed area. This is compared with the red channel profile extracted from the RGB map of the fluorescence image and presented in Figure 2 (b). The latter is used to identify the location of the tumour, healthy tissue, and the boundary between the two.

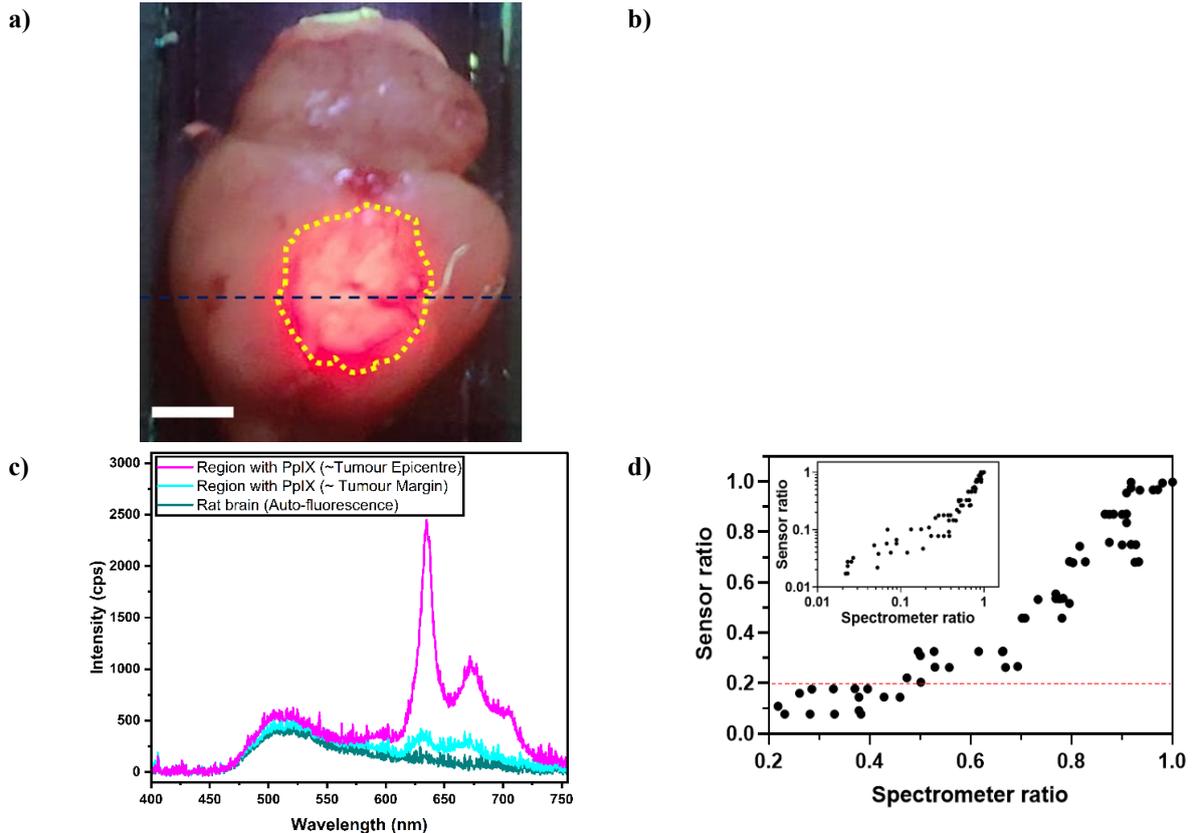

**Figure 2: (a) The fluorescence image of the rat brain. The yellow dotted line highlights the margin, blue dashed line indicating the scan direction – scale bar is 5 mm. (b) Plot of normalised ratio as a function of displacement obtained from the sensing system and compared to the fluorescence image data. (c) Fluorescence spectrum of the brain tissue at the centre of tumour region (pink), tumour margin (cyan) and healthy region (grey). The tumour margin is highlighted in (a). (d) The correlation between the**

**fluorescence ratio calculated using the spectrometer recordings and miniature sensing system at the tumour site – dotted line denotes the tumour margin. Insert: expanded to include healthy areas without PpIX emission (i.e., low ratios).**

As shown in Figure 2 (b), the tumour area identified from the image exhibits a flat top profile, while the miniature sensor produces a bell-shaped profile. The peak location of the sensor trace is at 11 mm displacement and matches the centre of the tumour (also at 11 mm). Tumour margin can be defined as the location where the fluorescence intensity diminishes. The sensor trace does not show a clear demarcation between the tumour and healthy tissue regions. Conversely, the image profile exhibits a clear difference between the two. We define the tumour margin as the location with a ratio of 0.9 on the image trace in Figure 2 (b). This point corresponds to the ratio of 0.21 +/- 0.03 value on the sensor trace. Although the sensor data does not exhibit an abrupt transition between the healthy and tumour, nevertheless, using the ratio of 0.21 sensor measurement as the demarcation point facilitates this distinction. It should be noted that the ratios reported here have been normalised to ease comparison.

Figure 2 (c) shows the fluorescence spectrum of the tissue acquired using a spectrometer. The healthy rat brain exhibits a fluorescence with peak wavelengths of 510 nm. Additional fluorescence peaks can be observed at the centre of the brain tissue with added tumour mimicking PpIX. They are centred at 635 nm for a primary emission and a secondary emission at 704 nm. At the region's boundary, similar peaks can be observed, although the PpIX related peaks are less pronounced. These fluorescence features are consistent with 5-ALA induced endogenous PpIX in brain tumours (Poulon et al. 2017).

Figure 2 (d) shows the correlation between the ratio calculated using the spectrometer recordings and the miniature sensing system. The ratios have been normalised to ease comparison. As discussed above, in this work, we have identified a sensor ratio of 0.21 as the tumour margin. Within the 0.21 to 1 ratio range (i.e., the tumour) there is a clear correlation between the sensor and spectrometer measurement. Indeed, results of the Spearman correlation indicated that there is a significant large positive correlation between the sensing system recording and spectrometer recordings ($r_s = 0.91$, $r_p < 0.001$). In the healthy brain region, where there is minimal emission from the PpIX, there is a poor correlation between the sensing system recording and the spectrometer recording ($r_s = 0.59$, $r_p < 0.001$). Brain tissue is a complex optical medium, and beyond the fluorescence, scattering and absorption contribute to the overall spectral measurement. The measurement was repeated on an optical phantom to understand this limitation better. Notably, the phantom lacked absorption and scattering agents. As shown in SI Fig 5 (b), there is a clear correlation between the optical sensor results and the spectrometer results, in the presence of PpIX. However, the correlation between the two measurements is poor in the healthy brain phantom. This is expected given that the $I_{635}$ in equation 2 approaches the noise level. It should be noted that the 0.21 threshold used in this work corresponds to a spectrometer ratio of 0.5. This is larger than the ratio value of ~ 0.3 identified in earlier work (2.1/6 margin to centre ratio) (Haj-Hosseini et al. 2010). Nevertheless, as shown in Figure 2 (b), in both cases the peak values coincide, and 0.21 thresholds are within 1 mm of each other.

## 3.2. Selectivity and specificity

Using the same arrangement reported in section 3.1, the red and green emissions from the brain tissue were measured in the lateral plane at a 1 mm step size. Based on the analysis in section 3.1, 20% of intensity was used to demarcate between the tumour and healthy tissue. This is illustrated in Figure 3 (a), with the white squares representing locations exceeding 20% of peak intensity. The results were compared with the actual tumour location. The true tumour location was identified by segmenting the image using the red channel threshold method at 90% of peak intensity. The composite image in Figure 3 (b) overlays the true tumour region over the image of the brain.

There is a clear agreement between the tumour regions identified by the sensor in Figure 3 (a) and the true tumour region in Figure 3 (b). This is better illustrated by overlapping the two images as shown in Figure 3 (c). To evaluate the selectivity and specificity of the sensor, the tumour region identified by the sensor is compared with the true tumour region over area intervals of 1 mm$^2$. This is used to calculate the true positive rate (sensitivity) and true negative rate (specificity). The receiver operating characteristic (ROC) curve in Figure 3 (d) shows the sensitivity as a function of 1-specificity for different threshold values. The area under the ROC is 0.923, pointing to a good diagnostic capability. Based on the ROC analysis, the sensor exhibits a sensitivity of 92.3% and a specificity of 98.3% at its optimum operating threshold.

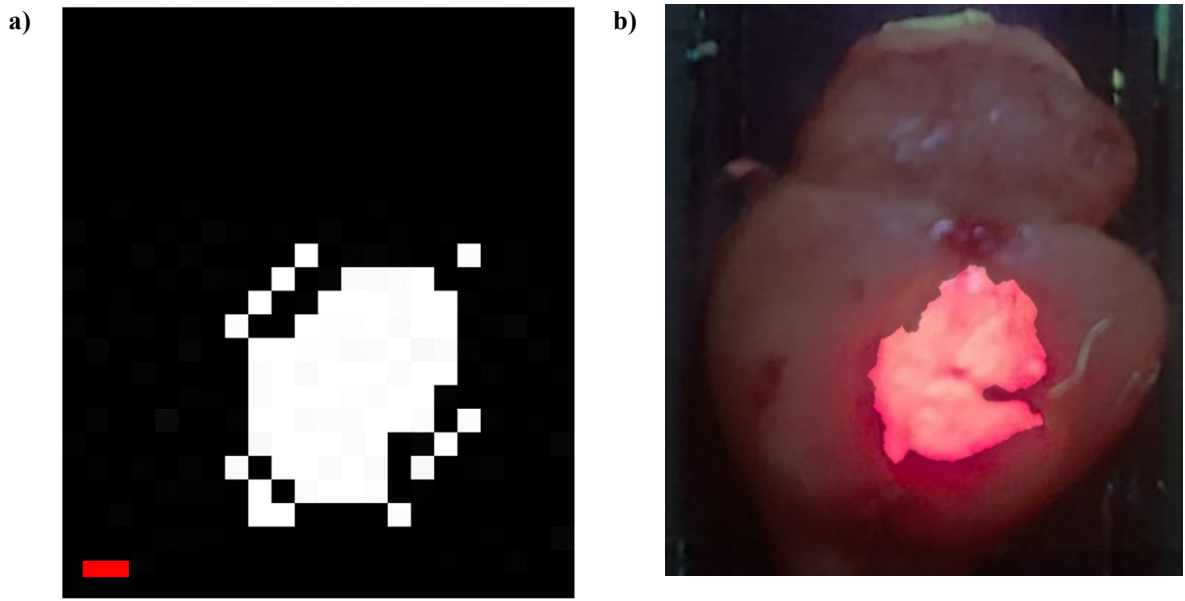

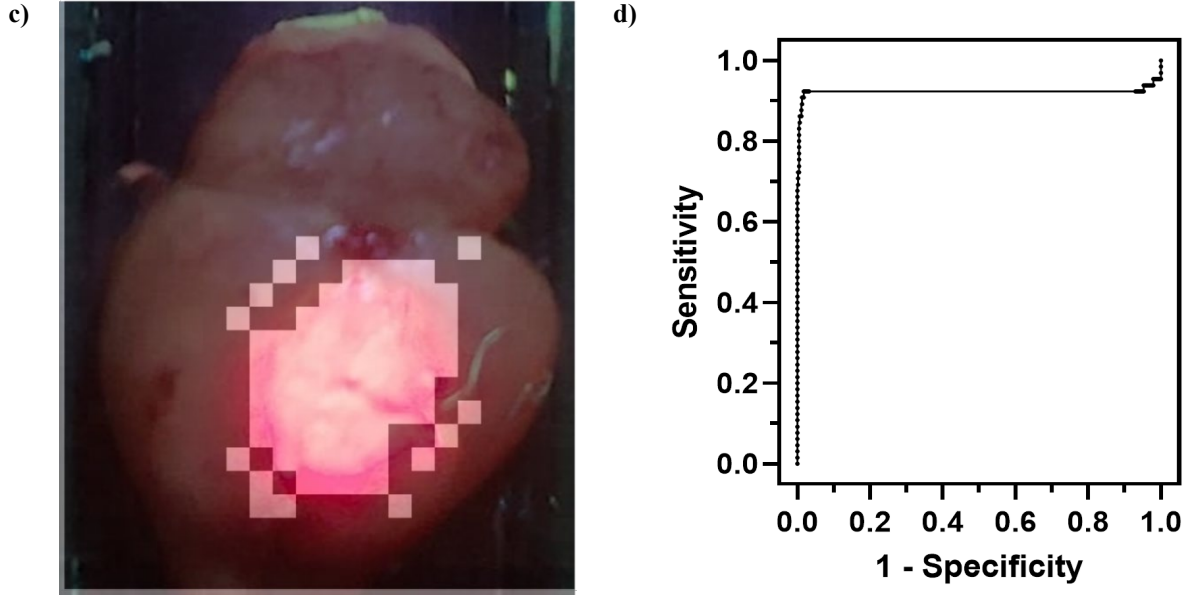

Figure 3: (a) Map of regions detected as tumour using the miniature sensing system. Each square is 1 mm × 1 mm, and the red scale bar is 2 mm. (b) composite image of the brain used in this study overlayed with tumour region detected by fluorescence imaging. (c) Composite image of the brain used in this study overlayed with tumour region detected by the miniature sensing system. (d) Receiver operating characteristic (ROC) curves comparing the outcomes detected by sensing system and fluorescence imaging at 1 mm × 1 mm resolution (area under the curve: 0.923, Sensitivity: 92.3%, 1-Specificity: 1.7%)

### 3.3. Contrast and intensity detection limits

The capability of the sensor chip to distinguish between the fluorescence emissions from the tumour and healthy tissue is critical. This is particularly the case when using a proposed miniature sensor chip that utilises filters instead of the conventional setups using spectrometers. The contrast detection study was carried out to investigate how discriminative the sensor chip is. In other words, when using a green light, we examined the intensity detected using the 514 nm sensor channel and compared it to the 635 nm sensor channel. Conversely, when using a red light, we measured how much we detected at the 635 nm sensor channel compared to the 514 nm channel. The measurement was performed using a green LED (peak: 518 nm) to emulate brain autofluorescence and a red LED (peak: 650 nm) to emulate tumour fluorescence. Table 1 shows the comparison between the sensor chip recording and spectrometer recording. As shown, the sensor chip has a 0.37% ~ 0.45% contrast detection error. In comparison, the contrast detection error of 1.33% ~ 1.73% was obtained when using the spectrometer.

Table 1: Detection selectivity of the system: Its capability in distinguishing between healthy and tumour fluorescence

| | SENSOR CHIP | |
|---|---|---|
| | ADC data (514 nm channel) | ADC data (635 nm channel) |
| Green light (Healthy tissue autofluorescence signature) | 57169 (99.63 %) | 212 (0.37 %) |
| Red light (Tumour fluorescence signature) | 279 (0.45 %) | 61653 (99.55 %) |
| SPECTROMETER | | |

|  | Spectral data (λ = 514 nm) | Spectral data (λ = 635 nm) |
|---|---|---|
| Green light (Healthy tissue autofluorescence signature) | 260552 (98.67 %) | 3499 (1.33 %) |
| Red light (Tumour fluorescence signature) | 4499 (1.73 %) | 255891 (98.27 %) |

The sensor chip's error can be attributed to the leakage of green light through the 635 nm sensor filter and the leakage of red light through the 514 nm channel filter, respectively. Indeed, these correspond to optical densities (OD) of 2.3 ~ 2.4, which are typical values for polymer-based filter technology designed for CMOS integration with the sensor chip. The fluorescence emitted from the brain tissue would have components in both red and green wavelengths. The red to green emission intensity ratio correlates with the MIB-1 index – an indicator of tumour cell proliferation activity (Kaneko and Kaneko 2016). This ratio can range from 0.7 for a tissue with an MIB-1 index of 0% to 30.9 for an MIB-1 index of 10% (Kaneko and Kaneko 2016). The typical ratio at the margin is reported as 1.3~6.5, with an MIB-1 index of 0.5%. Based on the values in Table 1, the 635 nm and 514 nm sensor channels have an error of 0.37% and 0.45%, respectively, leading to a red to green emission ratio error of 0.82%. This error is minimal compared to the 30.9 ~ 0.7 range observed in earlier surgical investigations, suggesting the newly developed setup could distinguish between low MIB-1 index at the tumour margin. Moreover, within the experimental error in this work, the measured values are consistent with the measurement obtained by the spectrometer.

The capability of the setup in detecting low fluorescence intensities was measured and is shown in SI Fig. 4. This is important as earlier works have linked the fluorescence intensities to the MIB-1 index (Kaneko and Kaneko 2016). Both 635 nm and 514 nm channels exhibited a noise equivalent power (NEP) of 4 nW/Hz$^{0.5}$. Ishihara et al. have reported emission-to-excitation (405 nm to 635nm) fluorescence intensity ratio in the range of 0.017 to 0.110 for tumours with MIB-1 labelling index of 0.3% to 22.6% (Ishihara et al. 2007). Based on a minimum ratio of 0.017, an excitation intensity of 0.235 µW or higher is needed to exceed the setup's NEP and detect the most challenging tumour. A typical excitation intensity used in fluorescence guided tumour resection surgery is 40 mW/cm$^2$ (Stummer et al. 1998), , corresponding to a power intensity of 4 µW over the detector area. This is over an order of magnitude over the minimum threshold of 0.235 µW., suggesting the newly developed setup could detect low fluorescence intensities generated by tumours with low MIB-1 labelling index.

As shown in SI Fig. 7 and Table 1, there is a clear separation between channels that ranges over two orders of magnitude, suggesting that the chip exhibits very high selectivity to the fluorescence signatures and a low error rate of less than 0.45%. Thus, it could measure the ratio in equation (2) and distinguish between healthy and tumour tissues.

### 3.4. Antifouling of diamond window

A key challenge in surgical optics is the fouling of the operative endoscopic window (Kreeft et al. 2017). As mentioned, any contact between the sensor surface and the body tissue or fluid may result in contamination of the endoscopic window, leading to unreliable results. This section investigates the effect of the contact between the window and a brain mass on the transmission intensities through the glass coverslip and a hydrogen-terminated diamond.

The analysis was achieved by measuring the changes in the transmission spectrum and total transmitted power across the glass and diamond windows as the results of brain tissue contact. A contact force of ~5 N over 5 seconds was selected. The spectral measurement is shown in Figure 4 a). Over ten measurement cycles, on average, the optical transmission through the diamond window reduced to 90% of the initial value at the PpIX fluorescence and autofluorescence emission wavelengths. This compares favourably with the average reduction of 60% across the glass sample at the same wavelengths. Moreover, the diamond window exhibited smaller variability in transmission (8%) across the ten measurement cycles compared to the glass window (17%). Moreover, as highlighted in Figure 4 (b), the glass window exhibits a larger variability compared to the diamond window for both green (autofluorescence) and red (PpIX channel). It should be noted that the initial transmission across the diamond sample is 68% and 75%, smaller than that of the glass sample, with 89% and 92% transmissions for the green and red channels, respectively. This difference can be attributed to the high refractive index of diamond, leading to considerable reflection at the air/diamond interfaces (Ahnood et al. 2020; Ahnood et al. 2016).

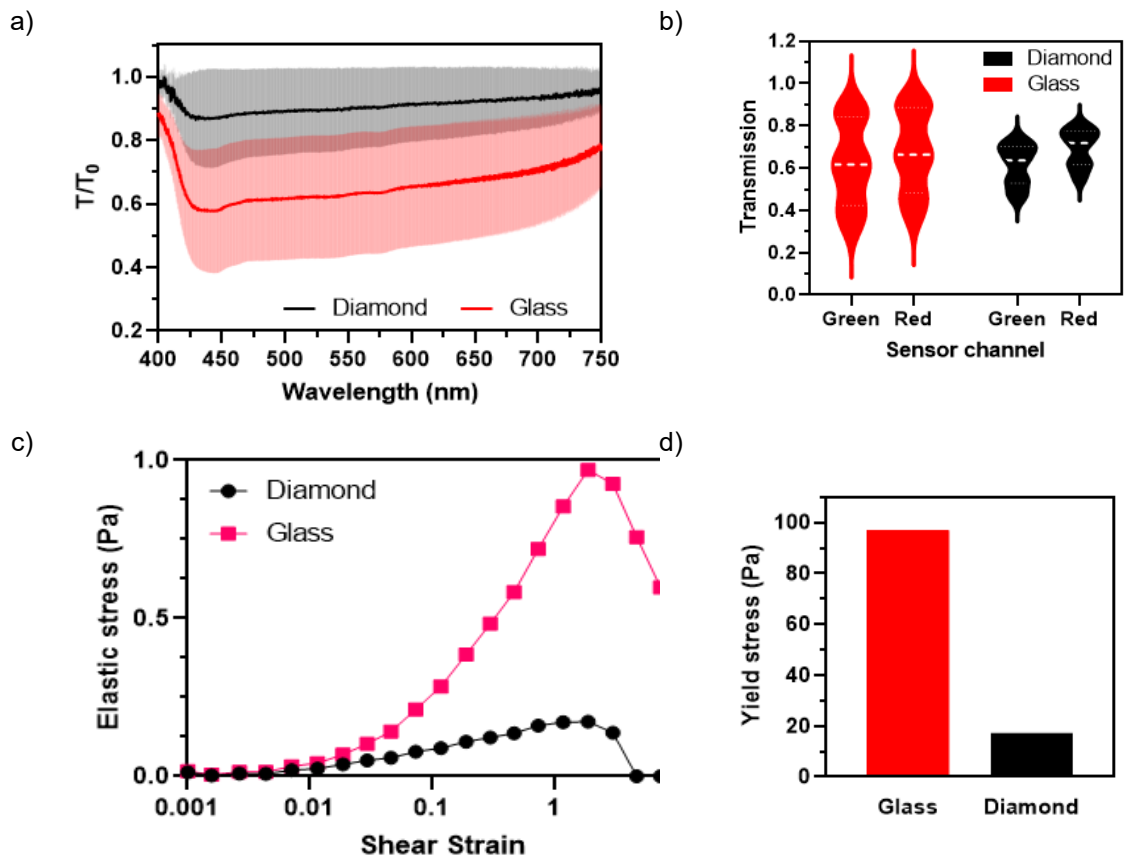

Figure 4: (a) Changes in the transmission spectrum through a diamond window compared to glass window as the result of tissue contact. T and $T_0$ denote transmission after and before contact with the tissue respectivly. The solid line represents the average change over 10 contact cycles, whilst the highlight area indicates the actual spectral changes. (b) Statistical distributions of transmission across the diamond and glass windows at the green and red sensor channels as the result of tissue contact. (c) The changes in the elastic stress due to the changes in the shear strain between the diamond/tissue

interface and glass/tissue interfaces. The maximum value in this stress profile corresponds to the yield point, i.e. the stress at which the brain-sample interface slips. (d) Yield stress of brain/glass and brain/diamond interfaces.

We quantified the relative adhesion strength of brain tissue to diamond and glass substrates using oscillatory rheology. In this measurement, the substrate of interest was mounted onto the rheometer, brought into contact with the tissue, and oscillated at increasing strain. Figure 4 (c) compares the change in elastic stress as a function of the shear strain between the diamond/tissue interface and glass/tissue interfaces. The yield stress of the tissue-substrate interface (i.e., the stress at which slippage occurs) is defined as the peak of the elastic stress curve (Walls et al. 2003; Yang et al. 1986). As shown in Figure 4 (d) As shown in Figure 4 (d) the yield stress of the glass/brain interface was 96 Pa, and that of the diamond substrate was 17 Pa. Thus, the diamond surface was almost six times less adhesive than the glass substrate.

4. Conclusion

In this work, we have investigated the viability of using a miniature fluorescence sensing system for tumour detection. The fluorescence sensing was performed directly at the site of the tumour surrogate. There was a strong correlation between the ratio measured by the sensing system and the conventional spectrometer, with a Spearman correlation coefficient of 0.91. Moreover, the sensing system provides a sufficiently low contrast error of less than 0.45%. These indicate that the proposed miniature fluorescence sensing system may be suitable for quantitative assessment of tumours – including ones with low MIB-1 index. Additionally, quantitative tumour assessment requires the capability of detecting low fluorescence intensities, as demonstrated in this work with an NEP of 4 nW/Hz$^{0.5}$. As well as more accurate quantitative tumour assessment, the latter opens the possibility of using an order of magnitude lower excitation intensity of 235 nW/mm$^2$ needed to minimise the PpIX photobleaching. Beyond quantitative tumour assessment, the miniature fluorescence system achieves a high accuracy in distinguishing between tumour and healthy tissues, with a sensitivity of 92.3% and specificity of 98.3% at 1 mm$^2$ spot size. Although the spot size could be further reduced through the integration of additional optical elements for focusing the light, this is sufficient to address a typical surgical margin (Haj-Hosseini et al. 2010; Santagata et al. 2014).

Light guided surgical resection tool requires a suitable optical window which minimises the collection of tissue derbies, blood, and fluids on its surface during the course of surgery. Here, we adopted a hydrophobic diamond window which exhibits a low-friction hydrophobic surface compared to a standard glass window. The diamond window exhibited a minimal change in its optical property (90% +/- 8%) compared to the glass window (60% +/- 17%). This along with other attributes of diamond, suggests that diamond window may be a suitable candidate for the application.

The miniature sensing approach demonstrated here allows direct detection of the light from the tumour. This eliminates the challenges in optical losses stemming from coupling light into optical fibres or collecting light using an imaging system. Moreover, the small size of the system introduced in this work may make it possible for its integration with tumour resection tools such as ultrasonic aspirators.

Acknowledgements

This work was supported by a CASS Foundation (Medicine and Science) Grant. The authors also appreciate the access granted to the facilities and the scientific and technical assistance of the Australian Microscopy & Microanalysis Research Facility at RMIT University.

Bibliography

Ahnood, A., Cheriton, R., Bruneau, A., Belcourt, J.A., Ndabakuranye, J.P., Lemaire, W., Hilkes, R., Fontaine, R., Cook, J.P., Hinzer, K., 2020. Laser Driven Miniature Diamond Implant for Wireless Retinal Prostheses. Advanced Biosystems 4(11), 2000055.

Ahnood, A., Fox, K.E., Apollo, N.V., Lohrmann, A., Garrett, D.J., Nayagam, D.A., Karle, T., Stacey, A., Abberton, K.M., Morrison, W.A., Blakers, A., Prawer, S., 2016. Diamond encapsulated photovoltaics for transdermal power delivery. Biosensors & bioelectronics 77, 589-597.

Al-Mefty, O., Kersh, J.E., Routh, A., Smith, R.R., 1990. The long-term side effects of radiation therapy for benign brain tumors in adults. Journal of neurosurgery 73(4), 502-512.

Belykh, E., Martirosyan, N.L., Yagmurlu, K., Miller, E.J., Eschbacher, J.M., Izadyyazdanabadi, M., Bardonova, L.A., Byvaltsev, V.A., Nakaji, P., Preul, M.C., 2016. Intraoperative fluorescence imaging for personalized brain tumor resection: current state and future directions. Frontiers in surgery 3, 55.

Belykh, E., Miller, E.J., Patel, A.A., Bozkurt, B., Yağmurlu, K., Robinson, T.R., Nakaji, P., Spetzler, R.F., Lawton, M.T., Nelson, L.Y., 2018. Optical characterization of neurosurgical operating microscopes: quantitative fluorescence and assessment of PpIX photobleaching. Sci Rep-Uk 8(1), 12543.

Bendsoe, N., Persson, L., Johansson, A., Axelsson, J., Svensson, J., Grafe, S., Trebst, T., Andersson-Engels, S., Svanberg, S., Svanberg, K., 2007. Fluorescence monitoring of a topically applied liposomal Temoporfin formulation and photodynamic therapy of nonpigmented skin malignancies. Journal of Environmental Pathology, Toxicology and Oncology 26(2).

Birriel, J.J., King, D., 2018. Fluorescence spectra of highlighter inks. The Physics Teacher 56(1), 20-23.

Bravo, J.J., Olson, J.D., Davis, S.C., Roberts, D.W., Paulsen, K.D., Kanick, S.C., 2017. Hyperspectral data processing improves PpIX contrast during fluorescence guided surgery of human brain tumors. Sci Rep-Uk 7(1), 9455.

Brown, P.D., Ballman, K.V., Cerhan, J.H., Anderson, S.K., Carrero, X.W., Whitton, A.C., Greenspoon, J., Parney, I.F., Laack, N.N., Ashman, J.B., 2017. Postoperative stereotactic radiosurgery compared with whole brain radiotherapy for resected metastatic brain disease (NCCTG N107C/CEC· 3): a multicentre, randomised, controlled, phase 3 trial. The Lancet Oncology 18(8), 1049-1060.

Brown, T.J., Brennan, M.C., Li, M., Church, E.W., Brandmeir, N.J., Rakszawski, K.L., Patel, A.S., Rizk, E.B., Suki, D., Sawaya, R., 2016. Association of the extent of resection with survival in glioblastoma: a systematic review and meta-analysis. JAMA oncology 2(11), 1460-1469.

Croce, A.C., Fiorani, S., Locatelli, D., Nano, R., Ceroni, M., Tancioni, F., Giombelli, E., Benericetti, E., Bottiroli, G., 2003. Diagnostic Potential of Autofluorescence for an Assisted Intraoperative Delineation of Glioblastoma Resection Margins¶. Photochemistry and Photobiology 77(3), 309-318.

Dandin, M., Abshire, P., Smela, E., 2007. Optical filtering technologies for integrated fluorescence sensors. Lab on a Chip 7(8), 955-977.

De Barros Bouchet, M.-I., Zilibotti, G., Matta, C., Righi, M.C., Vandenbulcke, L., Vacher, B., Martin, J.-M., 2012. Friction of diamond in the presence of water vapor and hydrogen gas. Coupling gas-phase lubrication and first-principles studies. The Journal of Physical Chemistry C 116(12), 6966-6972.

Duffau, H., 2005. Lessons from brain mapping in surgery for low-grade glioma: insights into associations between tumour and brain plasticity. The Lancet Neurology 4(8), 476-486.

Erkkilä, M.T., Bauer, B., Hecker-Denschlag, N., Madera Medina, M.J., Leitgeb, R.A., Unterhuber, A., Gesperger, J., Roetzer, T., Hauger, C., Drexler, W., 2019. Widefield fluorescence lifetime imaging of protoporphyrin IX for fluorescence-guided neurosurgery: An ex vivo feasibility study. Journal of Biophotonics 12(6), e201800378.

Guerrini, F., Roca, E., Spena, G., 2022. Supramarginal resection for glioblastoma: it is time to set boundaries! A critical review on a hot topic. Brain Sciences 12(5), 652.

Haj-Hosseini, N., Richter, J., Andersson-Engels, S., Wårdell, K., 2010. Optical touch pointer for fluorescence guided glioblastoma resection using 5-aminolevulinic acid. Lasers in Surgery and Medicine: The Official Journal of the American Society for Laser Medicine and Surgery 42(1), 9-14.

Hayward, I., 1991. Friction and wear properties of diamonds and diamond coatings. Surface and coatings technology 49(1-3), 554-559.

Ishihara, R., Katayama, Y., Watanabe, T., Yoshino, A., Fukushima, T., Sakatani, K., 2007. Quantitative spectroscopic analysis of 5-aminolevulinic acid-induced protoporphyrin IX fluorescence intensity in diffusely infiltrating astrocytomas. Neurologia medico-chirurgica 47(2), 53-57.

Johansson, A., Palte, G., Schnell, O., Tonn, J.C., Herms, J., Stepp, H., 2010. 5-Aminolevulinic acid-induced protoporphyrin IX levels in tissue of human malignant brain tumors. Photochemistry and photobiology 86(6), 1373-1378.

Kaibara, Y., Sugata, K., Tachiki, M., Umezawa, H., Kawarada, H., 2003. Control wettability of the hydrogen-terminated diamond surface and the oxidized diamond surface using an atomic force microscope. Diamond and Related Materials 12(3-7), 560-564.

Kaneko, S., Kaneko, S., 2016. Fluorescence-guided resection of malignant glioma with 5-ALA. International journal of biomedical imaging 2016.

Konicek, A.R., Grierson, D., Gilbert, P., Sawyer, W., Sumant, A., Carpick, R.W., 2008. Origin of ultralow friction and wear in ultrananocrystalline diamond. Physical review letters 100(23), 235502.

Kreeft, D., Arkenbout, E.A., Henselmans, P.W.J., Van Furth, W.R., Breedveld, P., 2017. Review of techniques to achieve optical surface cleanliness and their potential application to surgical endoscopes. Surgical Innovation 24(5), 509-527.

Kruijt, B., De Bruijn, H.S., Van Der Ploeg-van den Heuvel, A., De Bruin, R.W., Sterenborg, H.J., Amelink, A., Robinson, D.J., 2008. Monitoring ALA-induced PpIX photodynamic therapy in the rat esophagus using fluorescence and reflectance spectroscopy. Photochemistry and photobiology 84(6), 1515-1527.

Lakomkin, N., Hadjipanayis, C.G., 2019. The use of spectroscopy handheld tools in brain tumor surgery: current evidence and techniques. Frontiers in Surgery 6, 30.

Lauwerends, L.J., van Driel, P.B., de Jong, R.J.B., Hardillo, J.A., Koljenovic, S., Puppels, G., Mezzanotte, L., Löwik, C.W., Rosenthal, E.L., Vahrmeijer, A.L., 2021. Real-time fluorescence imaging in intraoperative decision making for cancer surgery. The Lancet Oncology 22(5), e186-e195.

Lehtonen, S.J., Vrzakova, H., Paterno, J.J., Puustinen, S., Bednarik, R., Hauta-Kasari, M., Haneishi, H., Immonen, A., Jääskeläinen, J.E., Kämäräinen, O.-P., 2022. Detection improvement of gliomas in hyperspectral imaging of protoporphyrin IX fluorescence–in vitro comparison of visual identification and machine thresholds. Cancer Treatment and Research Communications 32, 100615.

Li, M., Yang, T., Yang, Q., Wang, S., Fang, Z., Cheng, Y., Hou, X., Chen, F., 2023. Slippery quartz surfaces for anti-fouling optical windows. Droplet 2(1), e41.

Malinen, J., Rissanen, A., Saari, H., Karioja, P., Karppinen, M., Aalto, T., Tukkiniemi, K., 2014. Advances in miniature spectrometer and sensor development. Next-Generation Spectroscopic Technologies VII, pp. 83-97. SPIE.

Mehidine, H., Sibai, M., Poulon, F., Pallud, J., Varlet, P., Zanello, M., Devaux, B., Abi Haidar, D., 2019. Multimodal imaging to explore endogenous fluorescence of fresh and fixed human healthy and tumor brain tissues. Journal of Biophotonics 12(3), e201800178.

Mertens, M., Mohr, M., Bruehne, K., Fecht, H.-J., Łojkowski, M., Święszkowski, W., Łojkowski, W., 2016. Patterned hydrophobic and hydrophilic surfaces of ultra-smooth nanocrystalline diamond layers. Applied Surface Science 390, 526-530.

Nguyen, Q.T., Tsien, R.Y., 2013. Fluorescence-guided surgery with live molecular navigation—a new cutting edge. Nature reviews cancer 13(9), 653-662.

Poulon, F., Mehidine, H., Juchaux, M., Varlet, P., Devaux, B., Pallud, J., Abi Haidar, D., 2017. Optical properties, spectral, and lifetime measurements of central nervous system tumors in humans. Sci Rep-Uk 7(1), 13995.

Prickett, K.A., Ramsey, M.L., 2017. Mohs micrographic surgery.

Reichert, D., Erkkilae, M.T., Gesperger, J., Wadiura, L.I., Lang, A., Roetzer, T., Woehrer, A., Andreana, M., Unterhuber, A., Wilzbach, M., 2021. Fluorescence lifetime imaging and spectroscopic co-validation for protoporphyrin IX-guided tumor visualization in neurosurgery. Frontiers in Oncology 11, 741303.

Richter, J.C., Haj-Hosseini, N., Hallbeck, M., Wårdell, K., 2017. Combination of hand-held probe and microscopy for fluorescence guided surgery in the brain tumor marginal zone. Photodiagnosis and photodynamic therapy 18, 185-192.

Sanai, N., Berger, M.S., 2018. Surgical oncology for gliomas: the state of the art. Nature Reviews Clinical Oncology 15(2), 112-125.

Santagata, S., Eberlin, L.S., Norton, I., Calligaris, D., Feldman, D.R., Ide, J.L., Liu, X., Wiley, J.S., Vestal, M.L., Ramkissoon, S.H., 2014. Intraoperative mass spectrometry mapping of an onco-metabolite to guide brain tumor surgery. Proceedings of the National Academy of Sciences 111(30), 11121-11126.

Schupper, A.J., Rao, M., Mohammadi, N., Baron, R., Lee, J.Y., Acerbi, F., Hadjipanayis, C.G., 2021. Fluorescence-guided surgery: a review on timing and use in brain tumor surgery. Frontiers in Neurology 12, 682151.

Scott, M., Hopper, C., Sahota, A., Springett, R., McIlroy, B., Bown, S., MacRobert, A., 2000. Fluorescence photodiagnostics and photobleaching studies of cancerous lesions using ratio imaging and spectroscopic techniques. Lasers in medical science 15, 63-72.

Stummer, W., Pichlmeier, U., Meinel, T., Wiestler, O.D., Zanella, F., Reulen, H.-J., 2006. Fluorescence-guided surgery with 5-aminolevulinic acid for resection of malignant glioma: a randomised controlled multicentre phase III trial. The lancet oncology 7(5), 392-401.

Stummer, W., Stocker, S., Novotny, A., Heimann, A., Sauer, O., Kempski, O., Plesnila, N., Wietzorrek, J., Reulen, H., 1998. In vitro and in vivo porphyrin accumulation by C6 glioma cells after exposure to 5-aminolevulinic acid. Journal of Photochemistry and Photobiology B: Biology 45(2-3), 160-169.

Sunny, S., Cheng, G., Daniel, D., Lo, P., Ochoa, S., Howell, C., Vogel, N., Majid, A., Aizenberg, J., 2016. Transparent antifouling material for improved operative field visibility in endoscopy. Proceedings of the National Academy of Sciences 113(42), 11676-11681.

Surma-aho, O., Niemelä, M., Vilkki, J., Kouri, M., Brander, A., Salonen, O., Paetau, A., Kallio, M., J Pyykkönen, L., Jääskeläinen, J., 2001. Adverse long-term effects of brain radiotherapy in adult low-grade glioma patients. Neurology 56(10), 1285-1290.

Sutton, D., Limbert, G., Stewart, D., Wood, R., 2013. The friction of diamond-like carbon coatings in a water environment. Friction 1, 210-221.

Toulouse, A., Drozella, J., Thiele, S., Giessen, H., Herkommer, A., 2021. 3D-printed miniature spectrometer for the visible range with a 100× 100 µm 2 footprint. Light: Advanced Manufacturing 2(1), 20-30.

Valdes, P.A., Bekelis, K., Harris, B.T., Wilson, B.C., Leblond, F., Kim, A., Simmons, N.E., Erkmen, K., Paulsen, K.D., Roberts, D.W., 2014. 5-Aminolevulinic acid-induced protoporphyrin IX fluorescence in meningioma: qualitative and quantitative measurements in vivo. Neurosurgery 10(0 1), 74.

Valdes, P.A., Juvekar, P., Agar, N.Y., Gioux, S., Golby, A.J., 2019. Quantitative wide-field imaging techniques for fluorescence guided neurosurgery. Frontiers in Surgery 6, 31.

Valdes, P.A., Kim, A., Brantsch, M., Niu, C., Moses, Z.B., Tosteson, T.D., Wilson, B.C., Paulsen, K.D., Roberts, D.W., Harris, B.T., 2011. δ-aminolevulinic acid–induced protoporphyrin IX concentration correlates with histopathologic markers of malignancy in human gliomas: the need for quantitative fluorescence-guided resection to identify regions of increasing malignancy. Neuro-oncology 13(8), 846-856.

Valdés, P.A., Leblond, F., Kim, A., Harris, B.T., Wilson, B.C., Fan, X., Tosteson, T.D., Hartov, A., Ji, S., Erkmen, K., 2011. Quantitative fluorescence in intracranial tumor: implications for ALA-induced PpIX as an intraoperative biomarker. Journal of neurosurgery 115(1), 11-17.

Walls, H., Caines, S.B., Sanchez, A.M., Khan, S.A., 2003. Yield stress and wall slip phenomena in colloidal silica gels. Journal of Rheology 47(4), 847-868.

Wang, S.-W., Xia, C., Chen, X., Lu, W., Li, M., Wang, H., Zheng, W., Zhang, T., 2007. Concept of a high-resolution miniature spectrometer using an integrated filter array. Opt. Lett. 32(6), 632-634.

Wei, L., Roberts, D.W., Sanai, N., Liu, J.T., 2019. Visualization technologies for 5-ALA-based fluorescence-guided surgeries. Journal of Neuro-oncology 141, 495-505.

Yang, M.C., Scriven, L., Macosko, C., 1986. Some rheological measurements on magnetic iron oxide suspensions in silicone oil. Journal of Rheology 30(5), 1015-1029.

Yang, V.X., Muller, P.J., Herman, P., Wilson, B.C., 2003. A multispectral fluorescence imaging system: Design and initial clinical tests in intra-operative Photofrin-photodynamic therapy of brain tumors. Lasers in surgery and medicine 32(3), 224-232.

Zetterling, M., Elf, K., Semnic, R., Latini, F., Engström, E.R., 2020. Time course of neurological deficits after surgery for primary brain tumours. Acta neurochirurgica 162, 3005-3018.

Zhao, S., Wu, J., Wang, C., Liu, H., Dong, X., Shi, C., Shi, C., Liu, Y., Teng, L., Han, D., 2013. Intraoperative fluorescence-guided resection of high-grade malignant gliomas using 5-aminolevulinic acid–induced porphyrins: A systematic review and meta-analysis of prospective studies. PloS one 8(5), e63682.